%% file: main.tex
\documentclass[sigconf,screen]{acmart}

\usepackage{graphicx} 
\usepackage{cite}
\usepackage{hyperref}
\usepackage{float}
\usepackage{subfiles}
\usepackage{algorithm}
\usepackage{algpseudocode}
\usepackage{listings}
\usepackage{xcolor}
\usepackage{multirow}
\usepackage{multicol}
\usepackage{longtable}
\usepackage{booktabs}
\usepackage{subcaption}
\usepackage{graphicx}

\def\BibTeX{{\rm B\kern-.05em{\sc i\kern-.025em b}\kern-.08em
    T\kern-.1667em\lower.7ex\hbox{E}\kern-.125emX}}

\title{HEJ-Robust: A Robustness Benchmark for LLM-Based Automated Program Repair}

\author{Fazle Rabbi}
\orcid{0009-0007-8992-9682}
\affiliation{%
  \institution{Concordia University}
  \city{Montreal}
  \country{Canada}
}
\email{fazle.rabbi@mail.concordia.ca}

\author{Jinqiu Yang}
\orcid{0000-0003-4282-406X}
\affiliation{%
  \institution{Concordia University}
  \city{Montreal}
  \country{Canada}
}
\email{jinqiu.yang@concordia.ca}

\setcopyright{cc}
\setcctype{by}
\acmDOI{10.1145/3805760.3814929}
\acmYear{2026}
\copyrightyear{2026}
\acmISBN{979-8-4007-2601-9/2026/07}
\acmConference[AIware '26]{Proceedings of the 3rd ACM International Conference on AI-Powered Software}{July 6--7, 2026}{Montreal, QC, Canada}
\acmBooktitle{Proceedings of the 3rd ACM International Conference on AI-Powered Software (AIware '26), July 6--7, 2026, Montreal, QC, Canada}
\acmSubmissionID{fseaiware26main-pp011-data-p}
\received{2026-02-15}
\received[accepted]{2026-03-28}


\begin{document}

\begin{abstract}
Recent Large Language Models (LLMs) have shown strong performance on automated program repair across standard benchmarks. However, these benchmarks evaluate models on a single canonical form of buggy code and do not reflect the syntactic variations commonly observed in real-world software, leaving robustness largely unexamined. In this work, we construct HEJ-Robust, a robustness benchmark built from HumanEval-Java-Bug using eight semantics-preserving code transformations, resulting in 1,450 transformed instances. We evaluate five fine-tuned LLMs on this benchmark and show that model performance drops by over 50\% under several transformations, indicating that current LLM-based repair models lack robustness to minor syntactic variations.
\end{abstract}


\maketitle

\section{Introduction}
\label{introduction}
Automated program repair (APR) aims to automatically generate patches that fix buggy programs. Early APR approaches primarily followed the generate-and-validate paradigm, where candidate patches are synthesized using predefined or learned repair operators and validated against test suites. Representative systems include GenProg~\citep{GenProg}, PAR~\citep{kim2013automatic}, and systematic mutation-based repair techniques~\citep{qi2015analysis}.
While these approaches have demonstrated effectiveness on specific bug classes, they often suffer from scalability limitations and test-suite overfitting.

Recent advances in deep learning have significantly reshaped APR research by formulating program repair as a code translation problem, where buggy code is translated into its fixed version~\citep{jiang2021cure, zhu2021syntax}. Pre-trained LLMs, such as PLBART~\citep{ahmad2021unified}, and CodeT5~\citep{wang2021codet5}, have shown strong repair capability when fine-tuned on bug-fix data~\citep{zhang2022coditt5, chakraborty2022natgen, jiang2023impact}.
More recent studies further explore instruction-tuned and agent-based LLMs for automated program repair~\citep{fan2023automated, zhang2023critical, bouzenia2024repairagent}.
To evaluate these approaches, existing benchmarks commonly rely on Defects4J~\citep{just2014defects4j} or HumanEval-Java-Bug~\citep{jiang2023impact}, which assume a fixed syntactic representation of buggy programs.

Existing APR benchmarks evaluate repair accuracy on a single canonical buggy program, ignoring syntactic diversity among semantically equivalent code. Prior studies show neural code models are sensitive to semantics-preserving transformations~\citep{wang2023recode,rabbi2025multi}. While robustness testing via transformations, fuzzing, and adversarial examples has been studied in other SE tasks~\citep{pour2021search,yang2022natural}, robustness evaluation for LLM-based APR on function-level, human-crafted benchmarks remains largely unexplored. A related effort, Defects4J-TRANS~\citep{li2025evaluating}, applies transformations to project-level real bugs in Defects4J; our benchmark is complementary, focusing on function-level, human-crafted bugs from HumanEval-Java-Bug.

We address this gap by introducing a transformation-based robustness benchmark for automated program repair. Constructed by applying eight semantics-preserving transformations to HumanEval-Java-Bug~\citep{jiang2023impact}, our benchmark enables controlled evaluation of repair consistency. We use it to assess the robustness of five fine-tuned LLM repair models against code perturbations.

The contributions of this paper are as follows:
\begin{enumerate}
    \item We introduce a transformation-based robustness benchmark built on HumanEval-Java-Bug, covering eight semantics-preserving transformations and providing function-level, test-executable robustness evaluation for APR.
    \item We provide a systematic evaluation of LLM-based repair models under semantics-preserving transformations.
    \item We release the benchmark to facilitate future research on robust and reliable automated program repair.
\end{enumerate}

Our Code, dataset, and Artifacts are publicly available \footnote{https://github.com/frabbisw/hej-robust} 

\begin{table*}
    \caption{Fine-tuned models against different semantics-preserving code transformations on HumanEval-Java-Bug.}
    \label{results_finetuned}
    \centering
    \footnotesize
    \begin{subtable}{0.47\textwidth}
        \centering
        \caption{Local Variable Renaming (100 bugs)}
        \resizebox{\linewidth}{!}{%
            \begin{tabular}{crrrrr}
                \toprule
                ~ & \multicolumn{3}{c}{Pass@10} & \multicolumn{2}{c}{CodeBLEU} \\
                \cline{2-6}
                ~ & orig. & trans. & change &  orig.& trans. \\
                \midrule
                plbart\_base & 14.53 & 6.54 & 54.99$\downarrow$ & 82.11 & 81.91 \\
                plbart\_large & 21.88 & 9.91 & 54.71$\downarrow$ & 82.75 & 82.15 \\
                codet5\_small & 19.35 & 8.26 & 57.31$\downarrow$ & 82.17 & 81.43 \\
                codet5\_base & 24.81 & 12.28 & 50.5$\downarrow$ & 82.02 & 81.58 \\
                codet5\_large & 23.66 & 11.5 & 51.39$\downarrow$ & 80.74 & 80.92 \\
                \bottomrule
            \end{tabular}%
        }
    \end{subtable}
    \hspace{0.01\textwidth}
    \begin{subtable}{0.47\textwidth}
        \centering
        \caption{Method Renaming (149 bugs)}
        \resizebox{\linewidth}{!}{%
            \begin{tabular}{crrrrr}
                \toprule
                ~ & \multicolumn{3}{c}{Pass@10} & \multicolumn{2}{c}{CodeBLEU} \\
                \cline{2-6}
                ~ & orig. & trans. & change &  orig.& trans. \\
                \midrule
                plbart\_base & 19.46 & 18.13 & 6.83$\downarrow$ & 82.96 & 82.78 \\
                plbart\_large & 23.98 & 22.8 & 4.92$\downarrow$ & 83.34 & 83.04 \\
                codet5\_small & 21.16 & 19.46 & 8.03$\downarrow$ & 83.1 & 82.96 \\
                codet5\_base & 25.87 & 24.37 & 5.8$\downarrow$ & 83.06 & 82.71 \\
                codet5\_large & 24.75 & 23.98 & 3.11$\downarrow$ & 81.73 & 81.69 \\
                \bottomrule
            \end{tabular}%
        }
    \end{subtable}
    
    \vspace{0.5cm} 
    
    \begin{subtable}{0.47\textwidth}
        \centering
        \caption{Parameter Renaming (162 bugs)}
        \resizebox{\linewidth}{!}{%
            \begin{tabular}{crrrrr}
                \toprule
                ~ & \multicolumn{3}{c}{Pass@10} & \multicolumn{2}{c}{CodeBLEU} \\
                \cline{2-6}
                ~ & orig. & trans. & change &  orig.& trans. \\
                \midrule
                plbart\_base & 17.86 & 18.69 & 4.65$\uparrow$ & 83.63 & 83.84 \\
                plbart\_large & 22.6 & 23.33 & 3.23$\uparrow$ & 84.11 & 83.86 \\
                codet5\_small & 20.3 & 19.1 & 5.91$\downarrow$ & 83.92 & 84.02 \\
                codet5\_base & 24.77 & 24.41 & 1.45$\downarrow$ & 83.86 & 83.94 \\
                codet5\_large & 24.41 & 23.7 & 2.91$\downarrow$ & 82.67 & 82.55 \\
                \bottomrule
            \end{tabular}%
        }
    \end{subtable}
    \hspace{0.01\textwidth}
    \begin{subtable}{0.47\textwidth}
        \centering
        \caption{Boolean Exchange (7 bugs) $^\dagger$}
        \resizebox{\linewidth}{!}{%
            \begin{tabular}{crrrrr}
                \toprule
                ~ & \multicolumn{3}{c}{Pass@10} & \multicolumn{2}{c}{CodeBLEU} \\
                \cline{2-6}
                ~ & orig. & trans. & change &  orig.& trans. \\
                \midrule
                plbart\_base & 12.5 & 22.22 & 77.76$\uparrow$ & 86.03 & 85.8 \\
                plbart\_large & 22.22 & 30.0 & 35.01$\uparrow$ & 86.14 & 85.69 \\
                codet5\_small & 22.22 & 22.22 & 0\% & 83.68 & 83.3 \\
                codet5\_base & 22.22 & 12.5 & 43.74$\downarrow$ & 86.02 & 85.47 \\
                codet5\_large & 12.5 & 12.5 & 0\% & 85.26 & 84.67 \\
                \bottomrule
            \end{tabular}%
        }
    \end{subtable}
    
    \vspace{0.5cm}
    
    \begin{subtable}{0.47\textwidth}
        \centering
        \caption{Loop Exchange (142 bugs)}
        \resizebox{\linewidth}{!}{%
            \begin{tabular}{crrrrr}
                \toprule
                ~ & \multicolumn{3}{c}{Pass@10} & \multicolumn{2}{c}{CodeBLEU} \\
                \cline{2-6}
                ~ & orig. & trans. & change &  orig.& trans. \\
                \midrule
                plbart\_base & 19.32 & 18.39 & 4.81$\downarrow$ & 84.66 & 84.76 \\
                plbart\_large & 25.26 & 23.66 & 6.33$\downarrow$ & 84.77 & 85.5 \\
                codet5\_small & 21.55 & 17.92 & 16.84$\downarrow$ & 84.08 & 84.61 \\
                codet5\_base & 26.04 & 23.66 & 9.14$\downarrow$ & 84.58 & 85.12 \\
                codet5\_large & 28.28 & 26.8 & 5.23$\downarrow$ & 83.29 & 84.36 \\
                \bottomrule
            \end{tabular}%
        }
    \end{subtable}
    \hspace{0.01\textwidth}
    \begin{subtable}{0.47\textwidth}
        \centering
        \caption{Reorder Condition (603 bugs)}
        \resizebox{\linewidth}{!}{%
            \begin{tabular}{crrrrr}
                \toprule
                ~ & \multicolumn{3}{c}{Pass@10} & \multicolumn{2}{c}{CodeBLEU} \\
                \cline{2-6}
                ~ & orig. & trans. & change &  orig.& trans. \\
                \midrule
                plbart\_base & 16.88 & 15.69 & 7.05$\downarrow$ & 83.83 & 85.64 \\
                plbart\_large & 21.41 & 18.48 & 13.69$\downarrow$ & 84.17 & 86.1 \\
                codet5\_small & 19.7 & 17.92 & 9.04$\downarrow$ & 84.01 & 85.87 \\
                codet5\_base & 23.25 & 20.99 & 9.72$\downarrow$ & 83.94 & 85.75 \\
                codet5\_large & 23.45 & 21.62 & 7.8$\downarrow$ & 82.56 & 85.04 \\
                \bottomrule
            \end{tabular}%
        }
    \end{subtable}
    
    \vspace{0.5cm}
    
    \begin{subtable}{0.47\textwidth}
        \centering
        \caption{Insert Log Statement (173 bugs)}
        \resizebox{\linewidth}{!}{%
            \begin{tabular}{crrrrr}
                \toprule
                ~ & \multicolumn{3}{c}{Pass@10} & \multicolumn{2}{c}{CodeBLEU} \\
                \cline{2-6}
                ~ & orig. & trans. & change &  orig.& trans. \\
                \midrule
                plbart\_base & 17.22 & 16.43 & 4.59$\downarrow$ & 83.64 & 83.6 \\
                plbart\_large & 22.07 & 22.42 & 1.59$\uparrow$ & 84.1 & 83.85 \\
                codet5\_small & 19.53 & 18.4 & 5.79$\downarrow$ & 83.86 & 83.69 \\
                codet5\_base & 24.45 & 22.07 & 9.73$\downarrow$ & 83.8 & 83.57 \\
                codet5\_large & 24.78 & 24.45 & 1.33$\downarrow$ & 82.61 & 82.91 \\
                \bottomrule
            \end{tabular}%
        }
    \end{subtable}
    \hspace{0.01\textwidth}
    \begin{subtable}{0.47\textwidth}
        \centering
        \caption{Insert Try catch (114 bugs)}
        \resizebox{\linewidth}{!}{%
            \begin{tabular}{crrrrr}
                \toprule
                ~ & \multicolumn{3}{c}{Pass@10} & \multicolumn{2}{c}{CodeBLEU} \\
                \cline{2-6}
                ~ & orig. & trans. & change &  orig.& trans. \\
                \midrule
                plbart\_base & 16.91 & 13.74 & 18.75$\downarrow$ & 83.86 & 83.65 \\
                plbart\_large & 21.53 & 19.29 & 10.4$\downarrow$ & 84.17 & 83.94 \\
                codet5\_small & 19.29 & 11.02 & 42.87$\downarrow$ & 84.52 & 84.49 \\
                codet5\_base & 25.17 & 18.12 & 28.01$\downarrow$ & 84.29 & 84.3 \\
                codet5\_large & 26.14 & 18.12 & 30.68$\downarrow$ & 83.16 & 83.52 \\
                \bottomrule
            \end{tabular}%
        }
    \end{subtable}
\begin{minipage}{\textwidth}
\footnotesize $^\dagger$ Results for Boolean Exchange are based on only 7 instances and are statistically underpowered; they are excluded from headline robustness claims.
\end{minipage}
\end{table*}

\section{Related Work}
\label{related_work}
Automated program repair has been extensively studied over the past two decades. Early work primarily follows the generate-and-validate paradigm, where candidate patches are generated and validated against test suites~\citep{GenProg,qi2015analysis}. While effective on curated benchmarks such as Defects4J~\citep{just2014defects4j}, these approaches suffer from overfitting and scalability issues~\citep{yang2020exploring}.

More recently, deep learning-based APR approaches reformulate bug fixing as a neural machine translation problem, translating buggy code into fixed code~\citep{tufano2019empirical,lutellier2020coconut,jiang2021cure,zhu2021syntax}. Pre-trained LLMs further improve repair performance by leveraging large-scale code corpora before fine-tuning on repair data~\citep{wang2021codet5,ahmad2021unified,chakraborty2021multi,zhang2022coditt5,chakraborty2022natgen}. Most of these approaches evaluate on bug-fix pairs (BFPs)~\citep{tufano2019empirical,chakraborty2021multi}, which largely consist of abstract or canonicalized code. More recent benchmarks derived from HumanEval~\citep{chen2021evaluating} enable functional validation using test cases~\citep{jiang2023impact}. Complementary studies explore LLM-based repair in competitive programming and agent-based settings~\citep{fan2023automated,zhang2023critical,bouzenia2024repairagent}.

Parallel to APR research, robustness testing of neural models for code has gained attention. Prior work demonstrates that neural code models are vulnerable to small, semantics-preserving transformations~\citep{rabin2021generalizability}. Transformation-based testing, fuzzing, and adversarial example generation have been applied to code models~\citep{rabin2021generalizability,pour2021search}, with later work emphasizing natural and context-aware transformations~\citep{yang2022natural}. Works~\citep{wang2023recode, rabbi2025multi} evaluate the robustness of code generation models under semantics-preserving perturbations. More recently, Defects4J-TRANS~\citep{li2025evaluating} evaluates the generalizability of LLMs in APR by applying transformations to project-level bugs in Defects4J. Defects4J-TRANS applies five transformations: variable renaming, loop transformation, switch transformation, dead code injection, and boolean transformation. Our benchmark shares three of these (variable renaming, loop exchange, and boolean exchange), and adds five transformations not present in Defects4J-TRANS: method renaming, parameter renaming, condition reordering, log statement insertion, and try--catch insertion. Switch transformation was not applicable to HumanEval-Java-Bug as the dataset contains no switch statements. The two benchmarks are therefore complementary: Defects4J-TRANS targets project-level real bugs, while HEJ-Robust targets function-level, human-crafted bugs. While robustness has been studied for tasks such as code summarization and code representation learning, function-level robustness evaluation for automated program repair on human-crafted benchmarks remains largely unexplored. In particular, existing APR benchmarks do not systematically evaluate the robustness of repair models under semantics-preserving code transformations.

Beyond program repair, LLMs have been applied to related code tasks, including code translation~\citep{saha2024specification, rabbi2025babelcoder,rabbi2026beyond} and secure code generation~\citep{li2026exploratory,li2025secure,li2025prompt, cheng2025cfceval}, with studies also revealing reliability concerns such as social bias in LLM-generated code~\citep{ling2025bias, rabbi2026socialbias}. These findings collectively highlight that robustness and trustworthiness of LLM outputs remain open challenges across code tasks, motivating systematic evaluation frameworks such as the benchmark we propose.

\section{Robustness Benchmark Design}
\label{sec:Robustness}

\subsection{Base Dataset}
We adopt the HumanEval-Java-Bug dataset introduced by Jiang et al.~\citep{jiang2023impact}, which is derived from HumanEval~\citep{chen2021evaluating}. The dataset contains 164 Java programs with manually injected bugs and annotated buggy-line locations. Each instance is accompanied by executable test cases and human-written patches.
We select this dataset because it is manually curated, recent, and less likely to suffer from data leakage issues common in earlier APR benchmarks.

\subsection{Semantics-preserving Code Transformations}
\label{subsec:transformations}
We apply eight semantics-preserving code transformations that reflect common syntactic variations observed in real-world software. These transformations correspond to common coding practices such as identifier renaming during refactoring, loop restructuring for style conventions, and defensive programming patterns such as try-catch insertion and logging, making them representative of syntactic variation encountered in real-world Java development. The eight transformations are as follows:

\begin{enumerate}
    \item \textbf{Local variable renaming} renames all the occurrences of a local variable. An LLM generates the new identifier of the variable.
    \item \textbf{Method renaming} renames a method name using the same strategy as local variable renaming.
    \item \textbf{Parameter renaming} renames a parameter using the same strategy as local variable renaming and method renaming.
    \item \textbf{Log statement insertion} adds 
    \texttt{System.out.println("log")} as the first code statement in a method. Since HumanEval-Java-Bug is a function-level benchmark, test cases validate return values rather than standard output; this transformation therefore does not affect test outcomes.
    \item \textbf{Try-catch insertion} adds a \texttt{try-catch} block at 
    a random applicable code location.
    \item \textbf{Boolean exchange} changes the initialization value of a 
    boolean variable and wraps its subsequent uses in \texttt{!(...)} to 
    ensure semantic equivalence; for example, \texttt{boolean res = true;} becomes \texttt{boolean res = false;} and any subsequent \texttt{return res;} becomes \texttt{return !(res);}.
    \item \textbf{Loop exchange} replaces a \texttt{for} statement with an 
    equivalent \texttt{while} statement and vice versa.
    \item \textbf{Condition reordering} swaps the two operands of 
    \texttt{==} and \texttt{!=} expressions.
\end{enumerate}

\noindent Table~\ref{tab:examples} illustrates a before-and-after example for each of the eight transformations.

\begin{table}
\caption{Examples of the eight semantics-preserving transformations.}
\label{tab:examples}
\centering
\footnotesize
\begin{tabular}{p{0.18\linewidth} p{0.298\linewidth} p{0.37\linewidth}}
\toprule
\textbf{Transformation} & \textbf{Before} & \textbf{After} \\
\midrule
Local var.\ ren. 
  & \texttt{int temp = 0;} 
  & \texttt{int count = 0;} \\
\addlinespace
Method ren. 
  & \texttt{int compute()} 
  & \texttt{int calculate()} \\
\addlinespace
Parameter ren. 
  & \texttt{void foo(int x)} 
  & \texttt{void foo(int val)} \\
\addlinespace
Insert log 
  & \texttt{int foo(int x) \{} 
  & \texttt{int foo(int x) \{} \newline
    \texttt{~~System.out.println("log");} \\
\addlinespace
Insert try-catch 
  & \texttt{int r = f(s);} 
  & \texttt{try \{ int r = f(s); \}} \newline
    \texttt{catch (Exception e) \{\}} \\
\addlinespace
Boolean exc. 
  & \texttt{boolean res = true;} \newline \texttt{return res;}
  & \texttt{boolean res = false;} \newline \texttt{return !(res);} \\
\addlinespace
Loop exc. 
  & \texttt{for (int i=0;i<n;i++)} 
  & \texttt{int i=0;} \newline
    \texttt{while (i<n) \{...; i++;\}} \\
\addlinespace
Condition re. 
  & \texttt{if (a == b)} 
  & \texttt{if (b == a)} \\
\bottomrule
\end{tabular}
\end{table}

\noindent\textbf{Renaming transformations (1--3).}
For identifier renaming, we adopt the naturalness-aware substitution strategy proposed by Yang et al.~\citep{yang2022natural}. Unlike prior approaches that use random strings or fixed patterns~\citep{rabin2021generalizability,pour2021search}, this method generates context-aware and developer-natural identifiers, ensuring that performance degradation reflects robustness issues rather than unnatural code artifacts.

We use masked language prediction with CodeBERT and GraphCodeBERT to generate candidate identifiers and select substitutions based on cosine similarity in embedding space. Java code is parsed using tree-sitter~\citep{tree-sitter} to ensure consistent replacement across all occurrences. To control transformation strength, only one identifier is renamed per program.

\noindent\textbf{Structural and syntactic transformations (4--8).}
The remaining transformations are implemented using JavaTransformer~\citep{rabin2019testing}, which applies AST-based modifications via JavaParser. Transformations are applied only when syntactically valid. 

\noindent This yields 1,450 transformed instances in total.

\subsection{Benchmark Construction and Task Formulation}
\label{subsec:benchmark}
After applying transformations, the locations of buggy lines may change. We manually re-annotate the buggy-line locations for all transformed programs by inspecting each transformed instance and mapping the original buggy statement to its updated position. Instances where a transformation directly modifies the buggy line itself are removed from the dataset to avoid ambiguity. To reduce annotation errors, a second author independently verified more than 10\% of the re-annotated instances, and disagreements were resolved by discussion. Combined with the original human-written patches and test cases, this yields a fully executable benchmark suitable for robustness evaluation.
Model outputs are evaluated using both code-similarity metrics, such as CodeBLEU~\citep{ren2020codebleu,papineni2002bleu}, and functional correctness via test-based metrics (e.g., pass@10) provided by HumanEval-Java-Bug~\citep{jiang2023impact}.

The benchmark is extensible: new transformations can be added via JavaTransformer and re-annotated following the same protocol.

\section{Experimental Setup} To evaluate the proposed benchmark, we consider five LLMs: two PLBART variants (base and large) and three CodeT5 variants (small, base, and large), all fine-tuned and released by Jiang et al.~\citep{jiang2023impact}. We select these models because they are the only publicly available fine-tuned APR models evaluated on HumanEval-Java-Bug, which allows us to study robustness of APR-specific models under our transformations without introducing confounds from retraining. We directly evaluate these models without any modification to their original training or decoding configuration. Following the evaluation protocol of Jiang et al.~\citep{jiang2023impact}, we generate 10 candidate patches per bug and evaluate using Pass@10, where a bug is considered fixed if at least one generated patch passes all developer-written test cases. We do not modify seed settings, and we use the same generation procedure as provided in the original released models. 

Pass@10 serves as the primary metric for all robustness conclusions. CodeBLEU is reported as a reference control metric only; as our results confirm, it does not reliably capture functional robustness degradation. 

\balance

\section{Results}
\label{results}
The evaluation results are summarized in Table~\ref{results_finetuned}, which reports the performance of five fine-tuned models across eight transformed datasets. Each transformation is presented in a separate subtable, showing Pass@10 and CodeBLEU scores for both the original and transformed datasets. We also report the relative change from the original to the transformed dataset, indicated by $\uparrow$ for improvements and $\downarrow$ for degradations.

Across all eight transformations, we observe drops in Pass@10 for most models, with the largest degradation occurring under the Local Variable Renaming transformation, where performance decreases by 50.5\% to 57.31\%. This is likely because fine-tuned models rely heavily on identifier patterns learned during training; renaming local variables introduces distribution shifts in token sequences that disrupt the model's ability to identify the buggy location and generate a correct patch. Similar behavior has been reported in code generation robustness studies~\citep{wang2023recode}. In contrast, transformations such as Parameter Renaming and Insert Log Statement cause smaller drops, suggesting that method-level context or appended logging code is less disruptive to the repair process. Structural transformations such as Loop Exchange cause moderate drops, as they change control-flow structure while preserving variable names. Notably, robustness does not correlate with model size: larger models often degrade more than their smaller counterparts (e.g., \texttt{CodeT5\_large} vs.\ \texttt{CodeT5\_base}, and \texttt{PLBART\_large} vs.\ \texttt{PLBART\_base}) across multiple transformations. Results for Boolean Exchange are excluded from headline claims due to only 7 applicable instances; a single prediction shift 
changes the percentage by over 12 points, making any reported change uninterpretable. The apparent improvement for \texttt{plbart\_base} (+77.76\%) should be interpreted as noise rather than a meaningful robustness signal.

Two transformations show marginal improvements for some models. Under parameter renaming, \texttt{plbart\_base} and \texttt{plbart\_la rge} improve by 4.65\% and 3.23\% respectively. Although parameter renaming uses the same naturalness-aware substitution strategy as local variable renaming, the two transformations affect different parts of the code and may interact differently with the model's learned repair patterns. We do not draw strong conclusions from these marginal improvements, as they may reflect noise or dataset-specific characteristics rather than a systematic robustness effect.

In contrast, CodeBLEU scores remain largely stable across transformations, with only minor increases or decreases. This confirms that CodeBLEU is an unreliable indicator of  robustness for APR: models can produce syntactically similar but functionally incorrect patches without any detectable drop in CodeBLEU. We recommend that future APR robustness benchmarks rely on execution-based metrics such as Pass@k.

\section{Threats to Validity}
Our benchmark is constructed entirely from HumanEval-Java-Bug, which consists of small, isolated function-level programs; findings may not generalize to project-level bugs involving 
inter-file dependencies. The eight transformations do not cover all possible syntactic variations, and not all transformations are applicable to every program, explaining the varying instance counts per transformation. For renaming, we rename only one 
identifier per program, which may introduce bias across instances with different numbers of variables. Finally, we evaluate models from only two fine-tuned LLM families, PLBART and CodeT5; results may not generalize to instruction-tuned or agent-based LLMs.

\section{Conclusion}
We present \textbf{HEJ-Robust}, a robustness benchmark of 1,450 bug instances constructed from HumanEval-Java-Bug using 8 semantics-preserving transformations. Evaluating five fine-tuned LLMs, we show that even minor syntactic variations cause consistent drops in Pass@10, revealing substantial robustness gaps in current repair models. Future work will study richer transformation spaces, project-level benchmarks with inter-file dependencies, evaluation of instruction-tuned and agent-based LLMs in zero-shot settings, and improved robustness-aware training and evaluation metrics.

\section*{Acknowledgement}
\label{acknowledgement}
\noindent This research was supported by the Fonds de recherche du Québec (Grant No.2024-NOVA346499)\citep{nova}, Natural Sciences and Engineering Research Council of Canada (NSERC) through the Alliance, Grant (Grant No.586838-23), the NSERC Discovery Grant (Grant No. RGPIN-2019-07007 and Grant No. DGECR-2019-00464), and NSERC CREATE Grant (Grant No.555406-2021). We gratefully acknowledge the support of all funding agencies.

\clearpage
\newpage

\input{main.bbl}

\end{document}